\documentclass[11pt]{article}
\usepackage{psfig,float}
\newcommand{\Opc}{{\cal O}(p^4)}

\begin{document}

\begin{center}
{\Large {\bf Meson Meson and Meson Baryon Interactions in a chiral
Non-perturbative Approach}}
\end{center}

\vspace{0.3cm}

\begin{center}
{\Large {E. Oset$^1$, J.A. Oller$^1$, J.R. Pelaez$^2$ and A. Ramos$^3$ }}
\end{center}

\vspace{0.3cm}

{\small {\it
$^1$ Departamento de F\'{\i}sica Te\'orica and IFIC, Centro Mixto Universidad
de Valencia - CSIC, 46100 Burjassot (Valencia), Spain.

$^2$  Stanford Linear Accelerator Center, Stanford University, Stanford,
California 94309.

$^3$ Departament d'Estructura i Constituents de la Materia, Universitat
de Barcelona, Diagonal 647, 08028 Barcelona, Spain.

}}

\vspace{0.3cm}

\begin{abstract}
{\small{ A qualitative account of the meson-meson and meson-baryon
interactions using chiral \mbox{Lagrangians} and the inverse amplitude
method in coupled channels is done. The method, imposing exact
unitarity, proves to be a very useful tool to extend the  information
contained in the chiral Lagrangians at energies beyond the realm
of applicability of chiral perturbation theory.}}
\end{abstract}

\vspace{1cm}

\section{Introduction}

The meson-meson interaction has been the key problem to test Chiral
Perturbation Theory ($\chi PT$), which has proved rather successful at low
energies \cite{GaLe,ChPT}. The underlying idea is that an expansion in
powers of the meson momenta converges at sufficiently low energy, which in
practice is $\sqrt{s} \leq 500$ MeV. However, the convergence at higher
energies becomes progressively worse. Even more, one of the peculiar
features of the meson-meson interaction is the presence of resonances like
the $f_0$, $a_0$ in the scalar sector and the $\rho, K^*$ or the $\phi$
in the vector channels. These resonances will show up in the $T$ matrix as
poles that cannot be obtained using standard $\chi PT$. Nevertheless, the
constraints imposed by chiral symmetry breaking are rather powerful and not
restricted to the region where $\chi PT$ is meant to converge \cite{Steele}.

Two independent approaches of non perturbative character have extended the
use of chiral Lagrangians to higher energies and have been rather
successful, reproducing important features of the meson-meson interaction
including several resonances. One of them \cite{truongdo,IAM}, based upon the
Inverse Amplitude Method (IAM), first suggested in \cite{truong}, makes use
of $\chi PT$ amplitudes at ${\Opc}$. Elastic unitarity is imposed and thus no
mixture of channels is allowed. Then, the coefficients of the ${\cal O}(p^4)$
Lagrangian are fitted to the data. The absence of coupled channels has
obvious limitations, but in channels predominantly elastic the IAM is 
successful and able to generate 
dynamically the $\rho$, $K^*$ and $\sigma$  resonances, and to reproduce $%
\pi\pi$ scattering in the (I,J)=(0,0), (1,1), (2,0) partial waves, as well
as in the (3/2,0),(1/2,1) and (1/2,0) channels of $\pi K$ scattering. The
results are very successful up to 1 GeV in all these channels except the (0,0),
where it only yields good results up to 700 MeV. The limitations of this
single channel approach become evident, for instance, in the $f_0(980)$ and $%
a_0(980)$ resonances (J=0 and I=0 and 1, respectively) which do not appear. 

The second approach dealt with the J=0 sector \cite{olset}. The input
consists of the ${\cal O}(p^2)$ Lagrangian, which is used as the source of a
potential between mesons. This potential enters in a set of coupled channel
 Bethe-Salpeter (BS)
equations,  which leads to the
scattering matrix. The method imposes unitarity in coupled channels; hence
it yields inelasticities when inelastic channels open up. Amazingly, the
approach has only one free parameter,  which is a cut-off that regularizes
the loop integrals of the BS equation. Such a method proves rather
successful since phase shifts and inelasticities are reproduced accurately
up  to 1200 MeV. The $f_0(980)$ and $a_0(980)$ resonances appear as poles of
the $T$ matrix for I = 0 and 1, respectively, and their widths and partial
decay widths are very well reproduced. In addition, one finds a pole when I
= 0 at $\sqrt{s} \simeq 500$ MeV with a width of around 400 MeV,
corresponding to the $\sigma$ meson, which was also found with similar
properties with the IAM \cite{IAM}.

In this talk we will report on the method  proposed in \cite{prlm}
with applications to the meson-meson interaction and $K^-p$ interaction.
It consists of a generalization to coupled channels of the inverse 
amplitude method and unifies the two methods discussed above.

\section{Unitary amplitude in coupled channels}

 We denote by $T_{I J}$ the partial wave amplitude with isospin $I$ and angular
momentum $J$. For each value of $I$ and $J$ one has a definite channel with 
several meson-meson states coupled to each other. In Table I, we have listed 
these states for the $J = 0, 1$ channels.

\begin{center}
 {\small{Table I: Physical states used in the different $I,J$ channels}}
\end{center}

\begin{center}
{\small 
\begin{tabular}{|c|c|c|c|c|c|}
\hline
 & I = 0 & I = 1/2 & I = 1 & I = 3/2 & I = 2\\
\hline
J = 0 & $\begin{array}{cc}
\pi & \pi\\
K  & \bar{K} \end{array}$ &  
$\begin{array}{cc}
K & \pi\\
K  & \eta \end{array}$
&  $\begin{array}{cc}
\pi & \eta\\
K  & \bar{K} \end{array}$& $K \pi$& $\pi \pi$ \\
\hline
J = 1 & 
$K \bar{K}$ &  
$\begin{array}{cc}
K & \pi\\
K  & \eta \end{array}$
&  $\begin{array}{cc}
\pi & \pi\\
K  & \bar{K} \end{array}$&  &  \\
\hline
\end{tabular}}
\end{center}

Hence, throughout the present work, $T_{IJ}$ will be either a $2 \times 2$
symmetric matrix when two states couple, or just a number when there is only
one state. In what follows we omit the $I$, $J$ labels  and use a matrix
formalism, which will be valid for the general case  of $n\times n$ matrices
corresponding to $n$ coupled states.

Unitarity in coupled channels implies

\begin{equation}
{\rm Im}\, T_{if} = T_{in} \, \sigma_{nn} \, T^*_{nf}  \label{Tunit}
\end{equation}

\noindent where $\sigma$ is a real diagonal matrix whose elements account
for the phase space of the two meson intermediate states $n$ which are
physically accessible. With our normalization $\sigma_{nn}$ is given by the 
imaginary part of the loop integral of two meson propagators in the $n$ 
state

\[
\sigma_{nn} (s) = {\rm Im}\, G_{nn}(s) = - \frac{k_n}{8 \pi \sqrt{s}} \theta
(s-(m_{1n}+m_{2n})^2) 
\]

\begin{equation}
G_{nn}(s) = i \int \frac{d^4 q}{(2 \pi)^4} \; \frac{1}{q^2 - m^2_{1 n} + i
\epsilon} \; \frac{1}{ (P - q)^2 - m^2_{2 n} + i \epsilon}  \label{G}
\end{equation}

\noindent where $k_n$ is the on-shell CM momentum of the meson in the
intermediate state $n$, P is the initial total four-momentum and $m_{1 n},
m_{2 n}$ the masses of the two mesons in the state $n$.

From eq.(\ref{Tunit}) we can extract $\sigma$ and express it, in matrix
form, as

\begin{equation}
{\rm Im}\, G = T^{- 1} \cdot {\rm Im}\, T \cdot T^{* - 1}   
= \frac{1}{2 i} T^{- 1}\cdot (T - T^*)\cdot T^{* - 1}  
= \frac{1}{2 i} (T^{- 1 *} - T^{- 1}) = - {\rm Im}\, T^{- 1}  \label{ImG}
\end{equation}

Hence,

\begin{equation}
T^{- 1} = {\rm Re}\, T^{- 1} - i {\rm Im}\, G  \, ; \; \, \, \,
T = [{\rm Re}\, T^{- 1} - i \, {\rm Im}\, G]^{- 1}  \label{Tinverse}
\end{equation}

This is a practical way to write the unitarity requirements of eq.(\ref
{Tunit}) which tells us that we only need to know ${\rm Re}\, T^{- 1}$ since 
${\rm Im}\, T^{-1}$ is given by the phase space of the intermediate physical
states.

The next point is to realize that the $T$ matrix has poles associated to
resonances, which implies that the standard perturbative evaluation of $\chi
PT$ will necessarily fail close to these poles. As a consequence, one might
try to exploit the expansion of $T^{- 1}$, which will have zeros at the
poles of T, and in principle does not present convergence problems around 
the poles of T. With this 
idea in mind let us expand $T^{-1}$ in powers of $p^2$ as one would do 
for $T$ using $\chi PT$:

\begin{equation}
T \simeq T_2 + T_4 + ... ;\; \, \, \,
T^{- 1} \simeq T_2^{- 1}\cdot [1 + T_4 \cdot T_2^{- 1} ...]^{- 1} \simeq
T_2^{- 1}\cdot [1 - T_4 \cdot T_2^{- 1} ...]  \label{TIChPT}
\end{equation}

Multiplying formaly by $T_2 \,  T_2^{-1}$ to the right and by $T_2^{-1} T_2$ 
to the left, eq.(\ref{Tinverse}) can be rewritten as

\begin{equation}
T = T_2\cdot [T_2 \cdot {\rm Re}\, T^{- 1} \cdot T_2 - i T_2 \cdot {\rm Im}%
\, G \cdot T_2]^{- 1} \cdot T_2  \label{preT}
\end{equation}

Now, using the expansion for $T^{-1}$ of eq.(\ref{TIChPT}) we find
$
T_2 \cdot {\rm Re}\, T^{- 1} \cdot T_2 \simeq T_2 - {\rm Re}\, T_4 + ...,
$
and recalling that
$
{\rm Im}\, T_4 = T_2 \cdot {\rm Im}\, G \cdot T_2,
$
we finally obtain, within the ${\cal O}(p^4)$ approximation

\begin{equation}
T = T_2 \cdot [T_2 - T_4]^{- 1} \cdot T_2  \label{GenIAM}
\end{equation}

Note, as it is clear from eq.(\ref{preT}), that what we are expanding is
actually $T_2 \cdot {\rm Re}\, T^{-1} \cdot T_2$ which is also convergent 
for low energy.

This equation is a generalization to multiple
coupled channels of the IAM of ref.\cite{truongdo,IAM}. It makes the method
more general and powerful and also allows to evaluate transition cross
sections as well as inelasticities.

It is now important to realize that eq.(\ref{GenIAM}) requires the complete
evaluation of $T_4$, which is rather involved when dealing with many
channels, as it is the case here. This has been done in \cite{fgue} for 
the $K \bar{K}$ and $\pi \pi$ channels reproducing in very good agreement the 
experimental data up to around $\sqrt{s} \simeq$1.2 GeV for the (I,J)=(0,0),
(1,1) and (2,0) channels generating the $\sigma$, $f_0(980)$ and $\rho$ 
resonances. Instead, we present a further approximation to eq.(\ref{GenIAM}) 
which turns out to be technically much simpler and rather accurate. In order to illustrate the 
steps leading to our final formula, let us make before another approximation. 
Let us assume that through a suitable cut-off we can approximate

\begin{equation}
{\rm Re}\, \, T_4 \simeq T_2 \cdot {\rm Re}\, \, G \cdot T_2  \label{ReT4}
\end{equation}

In such a case we go back to the former equations and immediately
write

\begin{equation}
T = [1 - T_2 \cdot G]^{- 1} \cdot T_2 \Longrightarrow
T = T_2 + T_2 \cdot G \cdot T 
\end{equation}

\noindent
which is a BS equation for the $T$ matrix, where $T_2$ plays the role of the
potential. This is actually the approach followed in ref. \cite{olset}.

As we have already commented, the approximation of eq.(\ref{ReT4}) leads to
excellent results in the scalar channels. However, 
 the generalization to $J\neq 0$ is not possible since basic
information contained in the ${\cal O}(p^4)$ chiral Lagrangian is missing in
eq.(\ref{ReT4}). The obvious solution is to add a term to eq.(\ref{ReT4})
such that 

\begin{equation}
{\rm Re}\, \, T_4 \simeq T_4^P + T_2 \cdot {\rm Re}\, \, G \cdot T_2
\end{equation}

\noindent
where $T_4^P$ is the polynomial tree level contribution coming from the $%
{\cal O}(p^4)$ Lagrangian, whose terms contain several free parameters,
usually denoted $L_i$. Within our approach, these coefficients will be
fitted to data and denoted by $\hat{L}_i$ since they do not have to coincide
with those used in $\chi PT$, as we shall see. Actually, the $L_i$
coefficients depend on a regularization scale ($\mu$). In our scheme this
scale dependence appears through the cut-off.

The difference between \cite{fgue} and eq. (10) is that in \cite{fgue}
tadpoles and loops ind the cross channels are evaluated explicitly at 
${\Opc}$ while here they are absorbed into the $\hat L_i$ coefficients, so then
the values of $L_i$ in both approaches are  somewhat different.

Using eqs.(11) and former equations, our final formula for the $T$
matrix is given by

\begin{equation}
T = T_2 \cdot [T_2 - T_4^P - T_2\cdot G \cdot T_2]^{- 1} \cdot T_2
\label{ourT}
\end{equation}

\section{Results and comparison with experiment.}

Detailed calculations are presented in \cite{JAOO} for the different
channels. So here we just show some selected results in fig. 1. They 
are obtained using a cut off for the three momentum integration variable, 
$q_{max} = 1.02 \, GeV$.

As one can see, the results obtained are in good agreement with experiment
up to about $1.2 \, GeV$. In addition one also obtains poles in all the
meson resonances below that energy, the $\sigma (500), f_0 (980),
a_0 (980), K (800)$ in the scalar sector ($J = 0)$ plus the
$\rho (770)$ and $K^*  (800)$ in $J = 1, I=1$. A pole in $J = 1, I = 0$
corresponding to an $SU (3)$ octet and close to the $\phi$ meson is also
obtained. Partial decay widths are also calculated and are in fair
agreement with experiment \cite{JAOO}. The values of the $\hat{L}_i$ parameters
are of the same order as those of $\chi P T $ for a scale corresponding to our 
cut off $q_{max}$, with some discrepancies in $L_5$ and $2 L_6 + L_8$, but
as mentioned, tadpoles and crossed loops are incorporated at ${\Opc}$ 
by means of changes in these coefficients.

In the calculation of ref. \cite{fgue}, where tadpoles and crossed
loops are explicitly evaluated, the agreement between the $\hat{L}_i$ 
and $L_i$ coefficients
is better.

\begin{figure}
\hbox{
\psfig{file=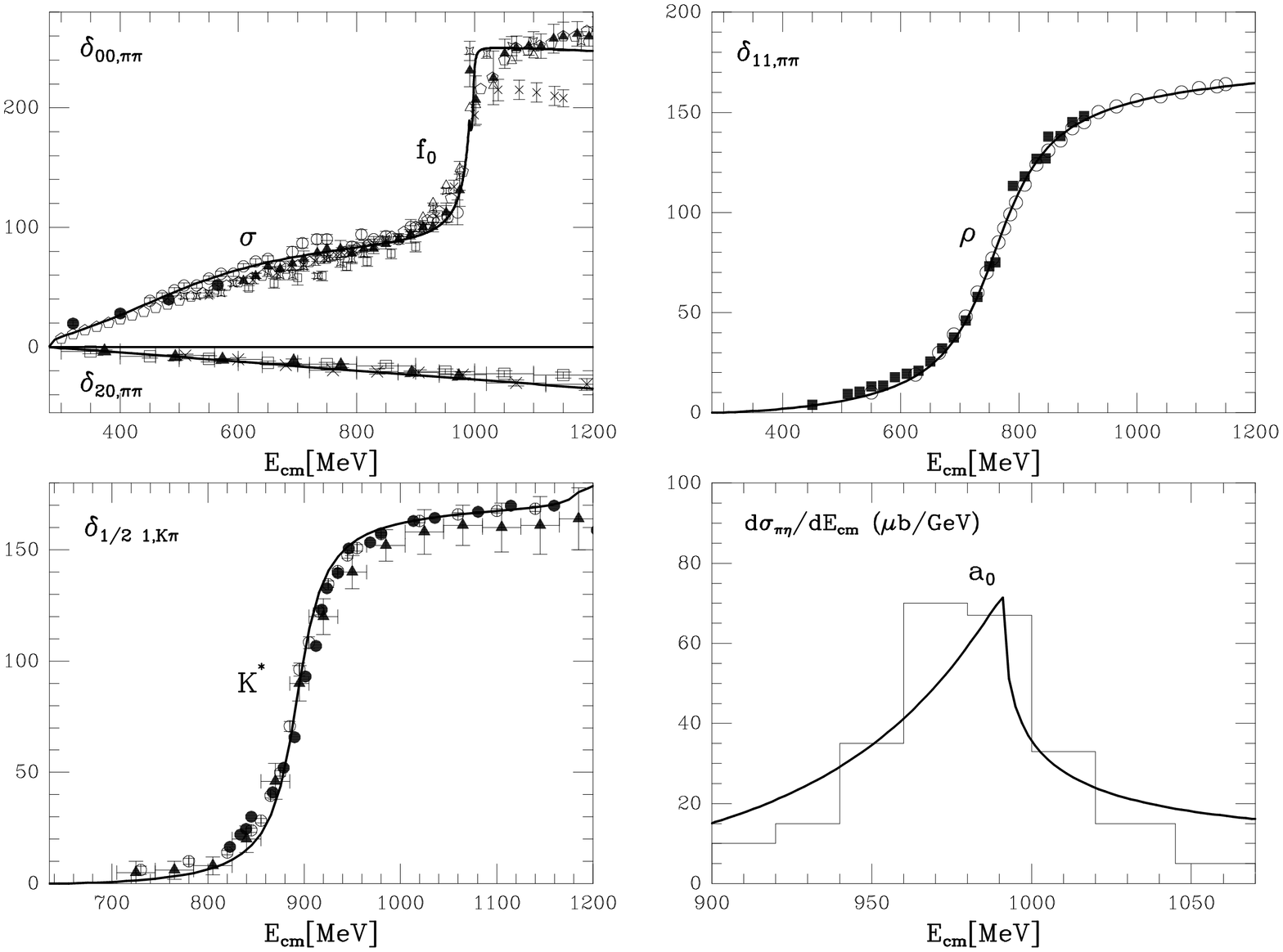,width=15cm,angle=-90}}
\hbox{
\psfig{file=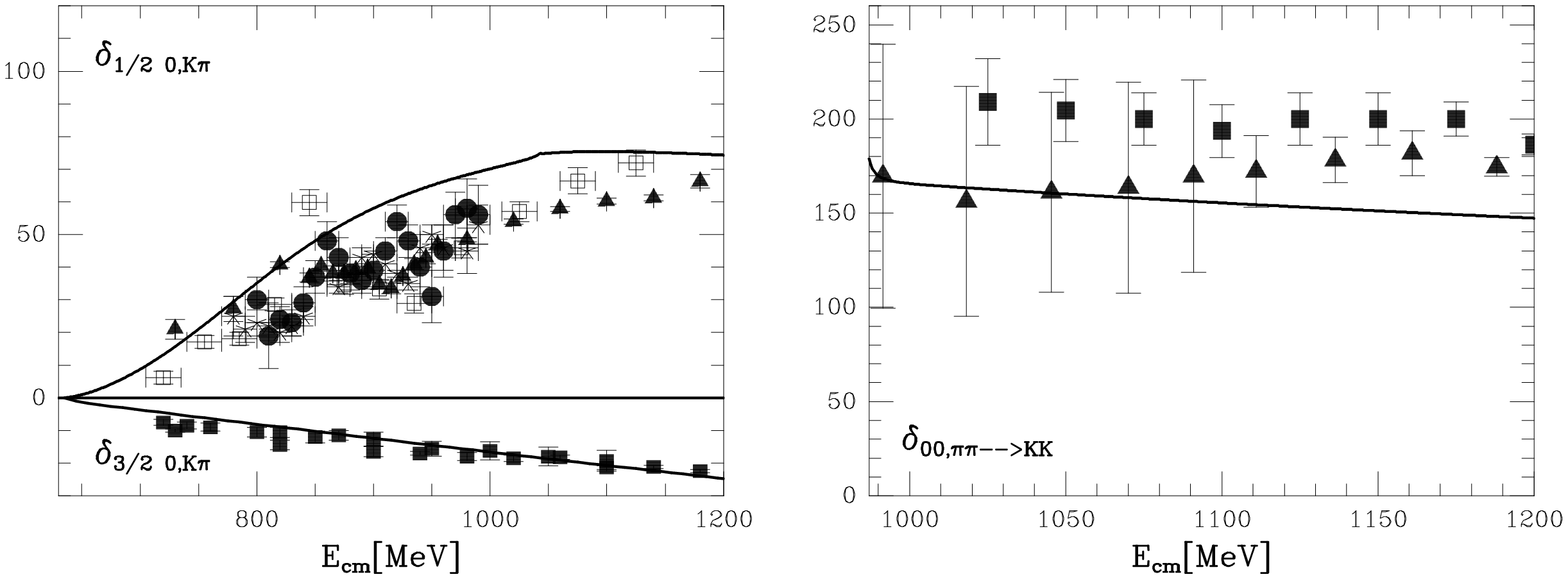,width=15cm,angle=-90}
}
{\bf Figure 1:}{We display the results of our method
for the phase shifts of $\pi\pi$
scattering in the $(I,J)=(0,0),(1,1),(2,0)$ channels, where the $\sigma$, 
$f_0$ and $\rho$ resonances appear, together with those of
$\pi\pi\rightarrow K \bar K$, as well as the phase shifts of $\pi K$
scattering in the $(3/2,0),(1/2,0) $ and $(1/2,1)$ channels, where we can see th
appearance of the $K^*$ resonance. The results also include
the $\pi^-\eta$ mass distribution 
for the $a_0$ resonance in the $(I,J)=(1,0)$ channel from
$K^-p\rightarrow \Sigma(1385)\pi^-\eta$.
For reference to the data, see \cite{truongdo} and \cite{olset}
and references therein.}
\end{figure}

\section{Coupled channel approach to s-wave $\bar{K} N$ interactions}

Here we follow the steps of the former section and include the
coupled chanels $K^- p, \bar{K}\,^0 n, \pi^0 \Lambda, \newline
\pi^+ \Sigma^-,
\pi^0 \Sigma^0, \pi^- \Sigma^+, \eta \Lambda, \eta \Sigma^0, K^+
\Xi^-, K^0 \Xi^0$ in order to study $K^- p$ elastic and inelastic
scattering close to threshold. The success of the approximation of 
eq. (8) for the $J = 0$ meson-meson interaction suggests this should be
sufficient here as it is indeed the case. Hence one uses the coupled
channel Bethe Salpeter eqns. of eq. (9) and uses the cut off
$q_{max}$ as a parameter. A value of $q_{max} = 630 \, MeV$ together
with a value for $f = 1.15 \, f_\pi$, between the $f_\pi$ and $f_K$,
was used in the calculations in \cite{EORA}. 

The lowest order $\chi PT$ amplitudes \cite{GaLe,ChPT} for these channels 
are easily evaluated and are given by

\begin{equation}
V_{i j} = - C_{i j} \frac{1}{4 f^2} \bar{u} (p') \gamma^{\mu} u
(p)
(k_{\mu} + k'_{\mu})
\end{equation}

\noindent
where $p, p' (k, k')$ are the initial, final momenta of the
baryons (mesons).
Also, for low energies one can safely neglect the spatial
components in eq.
(12) and only the $\gamma^0$ component becomes relevant, hence
simplifying
eq. (12) which becomes

\begin{equation}
V_{i j} = - C_{i j} \frac{1}{4 f^2} (k^0 + k'^0)
\end{equation}

\noindent
with  $C_{i j}$ a symmetric matrix which  is given in \cite{EORA}.

The scheme followed here is in the spirit of the one of refs. 
\cite{NKSI,NKTW}. The novelties here are the consideration of all the
meson channels in the coupled channel approach, while in \cite{NKSI,NKTW}
only six channels were considered, omitting the $\eta$ and $\Xi$
channels. In addition, a careful treatment of the renormalization of the 
lowest order constants when solving the scattering equations is done. While the $\Xi$ 
channels are of no practical relevance, the $\eta$ channels are important and 
change some cross sections by about a factor three. The results presented in 
\cite{NKSI,NKTW} are very similar to those obtained in \cite{EORA} 
because higher order terms in the chiral Lagrangians are included in 
\cite{NKSI,NKTW} by fitting some parameters and the effects of the $\eta$ 
channels are thus phenomenologically included.

The results obtained in \cite{EORA}, which are in good agreement with
data, are elastic $K^- p$ cross section, $K^- p \rightarrow \bar{K}\,^0 \pi,
\pi^0 \Lambda, \pi^+ \Sigma^-, \pi^0 \Sigma^0, \pi^- \Sigma^+ cross 
sections, K^- p$
and $K^- n$ scattering lengths, the $\Lambda (1405)$ resonance, 
which is generated dynamically, plus the threshold ratios
$\gamma = \Gamma (K^- p \rightarrow \pi^+ \Sigma^-)/ \Gamma (K^- p
\rightarrow \pi^- \Sigma^+)$,
$R_c = \Gamma (K^- p \rightarrow$ charged particles)/$\Gamma (K^- p
\rightarrow $ all), $R_n = \Gamma (K^- p \rightarrow \pi^0 \Lambda) /
\Gamma (K^- p
\rightarrow $ all neutral states).

\begin{figure}
\hbox{
\psfig{file=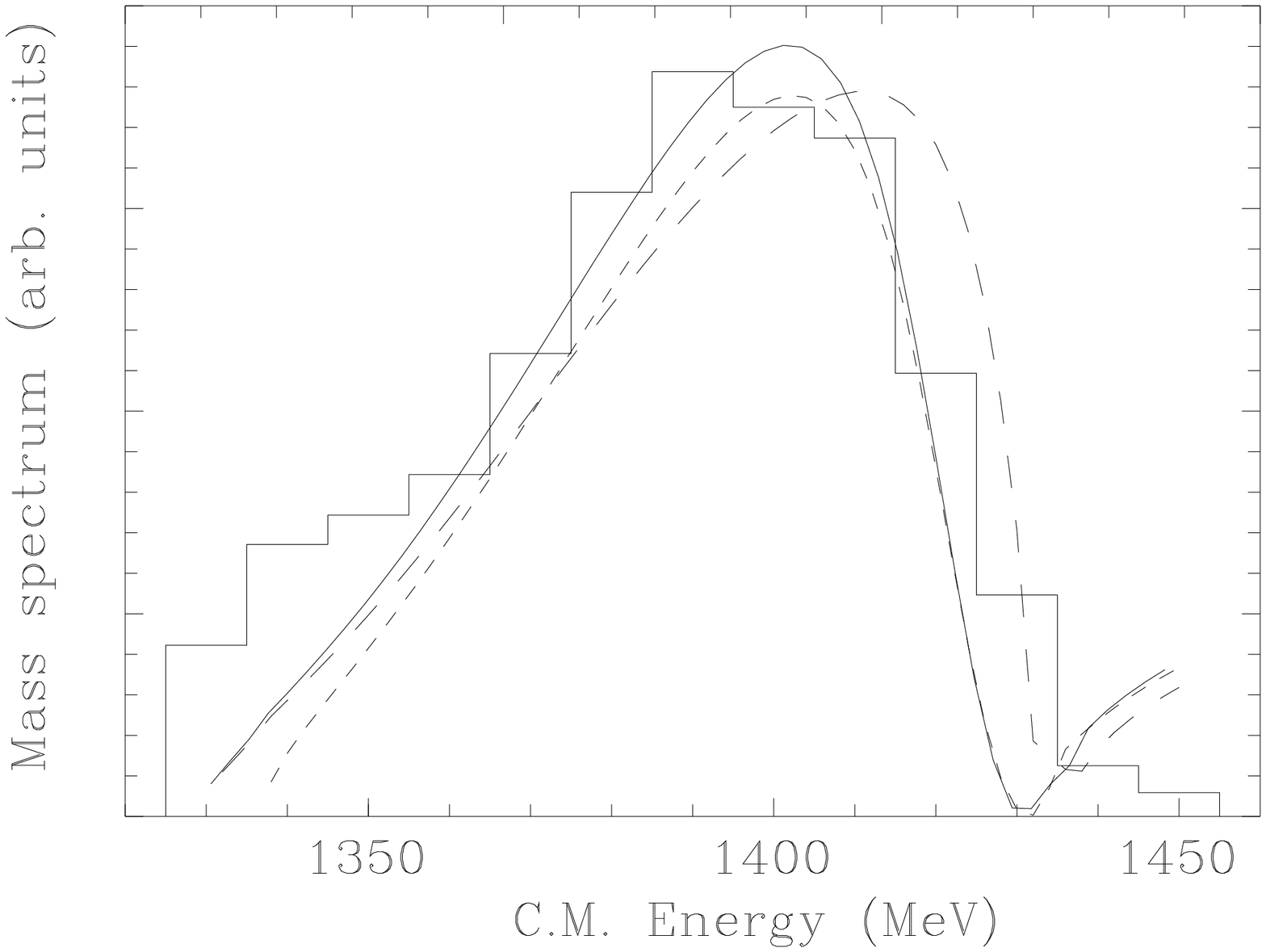,width=8.5cm,angle=-90}}
\hbox{
\psfig{file=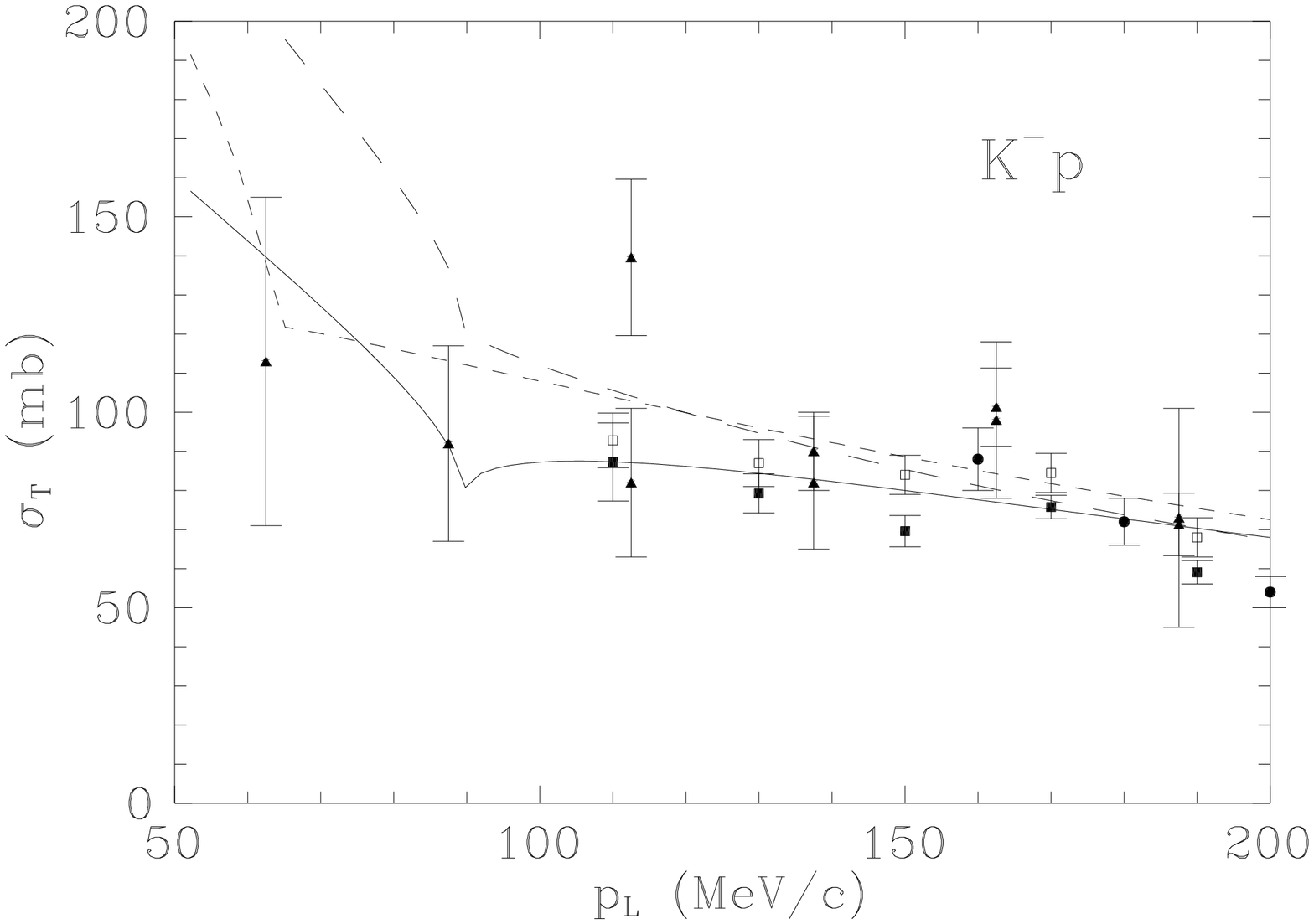,width=8.5cm,angle=-90}
}
{\bf Figure 1:}{From up to down. Mass spectra of $\pi \Sigma$ production 
corresponding to the $\Lambda$(1405) resonance. Elastic cross section for 
$K^-p$ collisions at low energies. The solid lines are the final results.}
\end{figure}

In fig. 2 we show the results obtained for the $\pi \Sigma$ mass 
distribution around the $\Lambda (1405)$ resonance plus the elastic 
$K^-p$ cross section. The quality of the agreement with data in the 
other channels is similar.

\section{Conclusions.}

We have shown how the inverse amplitude method in coupled channels,
respecting unitarity, allows one to go to higher energies, extracting
more information contained in the chiral Lagrangians than is possible
using $\chi P T$.

The description of the meson meson data upto 1.2 $GeV$ requires the
use of the $O (p^2)$ and $O (p^4)$ chiral Lagrangians. However, it is
remartable to see that both for the meson-meson interaction and for the
$\bar{K} N$ interaction in $J = 0$, the use of the lowest order Lagrangian
and a suitable cut off is enough to reproduce the experimental results
with high accuray.

The results obtained here obviously allow one to tackle typical problems
of $\chi P T$ at higher energies. Examples of that are the $\gamma \gamma
\rightarrow MM$ reaction which has been worked out in \cite{JAOS}, the
$\phi \rightarrow \gamma K^0 \bar{K}$ decay worked out in \cite{JAOL}
and the $K^- p \rightarrow \gamma \Lambda, \gamma \Sigma^0$ worked out
in \cite{TSLE}. The good results obtained for these reactions suggest that
the nonperturbative chiral scheme developed is an ideal tool to extend
the ideas of $\chi P T$ to much higher energies than are possible 
with the perturbative scheme.


\vspace{0.5cm}

\begin{thebibliography}{99}
\bibitem{GaLe}  {\footnotesize J. Gasser and H. Leutwyler, Ann. of Phys. 158
, (1984) 142, Nucl. Phys.  B250 (1985) 465. }

\bibitem{ChPT}  {\footnotesize U.G. Meissner, Rep. Prog. Phys. 56 
(1993) 903.\newline
V. Bernard, N. Kaiser and U.G. Meissner, Int. Jour. Mod. Phys. E4 
(1995) 193.\newline 
A. Pich, Rep. Prog. Phys. 58 (1995) 563.\newline 
G. Ecker, Prog. Part. Nucl. Phys. 35 (1995) 1. }

\bibitem{Steele}  {\footnotesize J. V. Steele, H. Yamagishi and I. Zahed,
Nucl. Phys. A615 (1997) 305. }

\bibitem{truongdo}  {\footnotesize A. Dobado, M. J. Herrero and T. N.
Truong, Phys. Lett. B235 (1990 134); A. Dobado and J. R. Pel\'{a}ez,
Phys. Rev. D47, (1993) 4883. }

\bibitem{IAM}  {\footnotesize A. Dobado and J. R. Pel\'{a}ez, Phys. Rev. 
D56 (1997) 3057. }

\bibitem{truong}  {\footnotesize T. N. Truong, Phys. Rev. Lett. 61 
(1988) 2526; 67, 2260 (1991). }



\bibitem{olset}  {\footnotesize J. A. Oller and E. Oset, Nucl. Phys. A620 
(1997) 438. }


\bibitem{prlm}  {\footnotesize J. A. Oller, E. Oset and J. R. Pel\'{a}ez., 
Phys. Rev. Lett. 80 (1998) 3452.}

\bibitem{fgue} {\footnotesize F. Guerrero and J.A. Oller, hep-ph/9805334, 
submitted to Nuc. Phys. B.}

\bibitem{JAOO} {\footnotesize J.A. Oller, E. Oset and J.R. Pelaez, 
hep-ph/9804209, submitted to Phys. Rev. D.}

\bibitem{EORA} {\footnotesize E. Oset and A. Ramos, nucl-th/9711022, Nucl. Phys. A}

\bibitem{NKSI} {\footnotesize N. Kaiser, P. B. Siegel and W. Weise, 
Nucl. Phys. A594, (1995) 325.}

\bibitem{NKTW} {\footnotesize N. Kaiser, T. Wass and W. Weise, Nucl. Phys. 
A612 (1997) 297.}

\bibitem{JAOS} {\footnotesize J.A. Oller and E. Oset,
 Nucl. Phys. A629 (1998) 739.}

\bibitem{JAOL} {\footnotesize  J.A. Oller, Phys. Lett. B 
in print , hep-ph/9803214}

\bibitem{TSLE} {\footnotesize
T.S.H. Lee, J.A. Oller, E. Oset and A. Ramos, nucl-th/9804053, submitted to 
Nucl. Phys. A.}

\end{thebibliography}
\end{document}